\documentclass[12pt]{article}
\usepackage{amsmath,amssymb,amsthm,bm,cite,fullpage,color,graphicx}

\title{The lowest excited configuration of harmonium}

\author{C.~L. Benavides-Riveros,$^{1,2}$
J.~M. Gracia-Bond\'ia$^{2,3}$
and J.~C. V\'arilly$^4$
\\ \\
$^1$Zentrum f\"ur Interdisziplin\"are Forschung, Wellenberg 1
\\
Bielefeld 33615, Germany
\\ \\
$^2$Departamento de F\'isica Te\'orica, Universidad de Zaragoza
\\ 
50009 Zaragoza, Spain
\\ \\
$^3$Instituto de F\'isica Te\'orica, CSIC--UAM, Madrid 28049, Spain
\\ \\
$^4$Escuela de Matem\'atica, Universidad de Costa Rica
\\
San Jos\'e 2060, Costa Rica}

\date{15 August 2012}

\newcommand{\al}{\alpha}              
\newcommand{\dl}{\delta}              
\newcommand{\Ga}{\Gamma}              
\newcommand{\ga}{\gamma}              
\newcommand{\la}{\lambda}             
\newcommand{\om}{\omega}              
\renewcommand{\th}{\theta}            
\newcommand{\vs}{\varsigma}           
\newcommand{\vth}{\vartheta}          
\newcommand{\even}{\mathrm{even}}     
\newcommand{\fs}{\mathrm{fs}}         
\newcommand{\gs}{\mathrm{gs}}         
\newcommand{\HF}{\mathrm{HF}}         
\newcommand{\odd}{\mathrm{odd}}       
\newcommand{\spin}{\mathrm{spin}}     

\DeclareMathOperator{\diag}{diag}     
\DeclareMathOperator{\Tr}{Tr}         

\newcommand{\vecform}{\bm}            
\newcommand{\PP}{\vecform{P}}         
\newcommand{\pp}{\vecform{p}}         
\newcommand{\RR}{\vecform{R}}         
\newcommand{\rr}{\vecform{r}}         
\newcommand{\xx}{\vecform{x}}         
\newcommand{\zz}{\vecform{z}}         

\newcommand{\dn}{{\mathord{\downarrow}}} 
\newcommand{\E}{\mathcal{E}}          
\newcommand{\half}{\tfrac{1}{2}}      
\newcommand{\N}{\mathbb{N}}           
\newcommand{\ox}{\otimes}             
\newcommand{\up}{{\mathord{\uparrow}}} 
\newcommand{\x}{\times}               
\let\dotover=\.                       
\renewcommand{\.}{\cdot}              

\newcommand{\dosFuno}[4]{{}_2F_1\biggl(\begin{matrix}
  #1,#2\\ #3\end{matrix}\,; #4\biggr)} 
\newcommand{\ketbra}[2]{|#1\rangle\langle#2|} 
\newcommand{\set}[1]{\{\,#1\,\}}    
\newcommand{\twobytwo}[4]{\begin{pmatrix}#1& #2\\ #3& #4\end{pmatrix}}
\newcommand{\word}[1]{\quad\mbox{#1}\quad} 

\makeatletter
\def\section{\@startsection{section}{1}{\z@}{-3.5ex plus -1ex minus
 -.2ex}{2.3ex plus .2ex}{\large\bfseries}}
\def\subsection{\@startsection{subsection}{2}{\z@}{-3.25ex plus -1ex
 minus -.2ex}{1.5ex plus .2ex}{\normalsize\bfseries}}
\makeatother

\begin{document}

\maketitle

\begin{abstract}
The harmonium model has long been regarded as an exactly solvable
laboratory bench for quantum chemistry \cite{Heisenberg26}. For
studying correlation energy, only the ground state of the system has
received consideration heretofore. This is a spin singlet state. In
this work we exhaustively study the lowest excited (spin triplet)
harmonium state, with the main purpose of revisiting the relation
between entanglement measures and correlation energy for this quite
different species. The task is made easier by working with Wigner
quasiprobabilities on phase space.
\end{abstract}


\section{Introduction}
\label{sec:introibo}

Replacing the wave function of electronic systems by the reduced
2-body density matrix~$\ga_2$ tremendously saves computation without
losing relevant physical information. Till very recently, the
solutions to the $N$-representability for that matrix
\cite{GarrodP64,AyersGL06} were impractical. This certainly did not
impede great advances in the use of~$\ga_2$ for many-electron quantum
systems ---see for instance~\cite{Mazziotti12a}. Now a constructive
solution \cite{Mazziotti12b} to that representability problem, leading
to a hierarchy of constraints \cite{Mazziotti12c} on the variation
space for $\ga_2$, has been unveiled.

At any rate, the last fifteen years have witnessed a justifiable
amount of work in trying to obtain the 2-body matrix as a functional
of the 1-body density matrix $\ga_1$. Starting with the pioneer work
by M\"uller \cite{Mueller84}, several competing functionals have been
designed, partly out of theoretical prejudice, partly with the aim of
improving predictions for particular systems. We shall discuss
\textit{pure} state representability for $\ga_1$ in the case of our
interest in Section \ref{sec:NWO}.

Two-electron systems are special in that $\ga_2$ is known ``almost
exactly'' in terms of $\ga_1$. Let us express $\ga_1$ by means of the
spectral theorem in terms of its natural orbitals and occupation
numbers. For instance, the ground state of the system admits a
1-density matrix:
\begin{equation}
\ga_1(\xx,\xx') = \bigl( \up_1\up_{1'} + \dn_1\dn_{1'} \bigr)
\ga_1(\rr,\rr') = \bigl( \up_1\up_{1'} + \dn_1\dn_{1'} \bigr)
\sum_i n_i \, \phi_i(\rr)\phi_i^*(\rr').
\label{eq:ad-porcos} 
\end{equation}
Here
$\sum_i n_i = 1$. Mathematically this a mixed state. The corresponding
2-density matrix is given by
\begin{align}
\ga_2(\xx_1,\xx_2;\xx'_1,\xx'_2)
&= \bigl( \up_1\dn_2 - \dn_1\up_2 \bigr)
\bigl( \up_{1'}\dn_{2'} - \dn_{1'}\up_{2'}) \sum_{ij} 
\frac{c_ic_j}{2}\, \phi_i(\rr_1) \phi_i(\rr_2) \phi_j^*(\rr'_1)
\phi_j^*(\rr'_2),
\notag \\
&\quad \word{with coefficients}  c_i = \pm \sqrt{n_i}. 
\label{eq:margaritas-vestras} 
\end{align}
The expression is exact, but the signs of the $c_i$ need to be
determined to find the ground state \cite{LoewdinS56,Kutzelnigg63}.
Note that $\ga^2_2 = \ga_2$. The first excited state of the system
admits a reduced 1-density matrix of the kind:
$$
\ga_1(\xx;\xx') = (\text{spin factor}) \x \sum_{ij} n_i \,
\bigl(\phi_{2i}(\rr) \phi^*_{2i}(\rr')
+ \phi_{2i+1}(\rr) \phi^*_{2i+1}(\rr') \bigr)
$$
with $\sum_i n_{i} = 1$ and $\spin \in \set{ \up_1\up_{1'}, \
\half(\up_1\up_{1'} + \dn_1\dn_{1'}), \ \dn_1\dn_{1'} }$. The
corresponding spinless \mbox{2-density} matrix
$\ga_2(\rr_1,\rr_2;\rr'_1,\rr'_2)$ is given by
\begin{align*}
\sum_{ij} \frac{c_ic_j}2
& \bigl[ \phi_{2i}(\rr_1) \phi_{2i+1}(\rr_2)
\phi_{2j}^*(\rr'_1) \phi_{2j+1}^*(\rr'_2)
+ \phi_{2i+1}(\rr_1) \phi_{2i}(\rr_2) \phi_{2j+1}^*(\rr'_1)
\phi_{2j}^*(\rr'_2)
\\[-\jot]
&\quad - \phi_{2i}(\rr_1) \phi_{2i+1}(\rr_2) \phi_{2j+1}^*(\rr'_1)
\phi_{2j}^*(\rr'_2)
- \phi_{2i+1}(\rr_1) \phi_{2i}(\rr_2) \phi_{2j}^*(\rr'_1)
\phi_{2j+1}^*(\rr'_2) \bigr],
\\[\jot]
&\qquad \word{with coefficients} c_i = +\sqrt{n_i}. 
\end{align*}
Due to the antisymmetry of this state, there is no ambiguity in the
choice of sign.

\medskip

A completely integrable analogue of a two-electron atom, here called
\textit{harmonium}, describes two fermions interacting with an
external harmonic potential and repelling each other by a Hooke-type
force; thus the harmonium Hamiltonian in Hartree-like units is
\begin{equation}
H = \frac{p_1^2}{2} + \frac{p_2^2}{2} + \frac{k}{2}(r_1^2 + r_2^2)
- \frac{\dl}{4} r^2_{12},
\label{eq:Mosh-atom} 
\end{equation}
where $r_{12} := |\rr_1 - \rr_2|$. This model is rooted in the history
of quantum mechanics: Heisenberg first invoked it to approach the
spectrum of helium~\cite{Heisenberg26}.

Several problems related with this model ---although not quite the
present one--- are analytically solved; and so it is tempting to
employ it as a testing ground for methods used in other systems, such
as the helium series. Indeed, Moshinsky \cite{Moshinsky68}
reintroduced it with the purpose of calibrating correlation energy.
There is considerable interest nowadays on learning from harmonium,
including further study of correlation
\cite{MarchCCA08,Loos10,NagyP11}, approximation of functionals
\cite{AmovilliM03,NagyP10}, and beyond quantum chemistry, questions of
entanglement \cite{AmovilliM04,PipekN09,YanyezPD10,BouvrieMPSMD12} and
black hole entropy~\cite{Srednicki93}.

In the past, harmonium problems have been attacked with ordinary wave
mechanics \cite{Davidson76}. Now, for the analysis of harmonium the
\textit{phase space} representation of quantum mechanics recommends
itself. The deep reason for this is the metaplectic invariance of that
formalism \cite{Iapetus}, hidden in the standard approach: this made
it easy to solve the sign dilemma in the exact
L\"owdin--Shull--Kutzelnigg formula~\cite{LoewdinS56,Kutzelnigg63} for
$\ga_2$ in terms of~$\ga_1$, for two-electron systems
\cite{Pluto,Pallene}. We come to this at the end of the next section.
Such a phase-space description was taken up first by
Dahl~\cite{Dahl09}, and then developed, within the context of a
phase-space density functional theory (WDFT), by Blanchard,
Ebrahimi-Fard and ourselves
\cite{Pluto,Pallene,Hermione,Bellona,Marmulla}.

Our goal in this article is to understand, in WDFT terms, the first
excited state of harmonium. As for helium-like atoms, we expect it to
be the lowest spin triplet state, to which we refer simply as the
triplet. Particularly we make clear the nonexistence of a phase
dilemma in this situation, and pinpoint the similarities and
differences between the relative behavior of entropy and correlation
energy for the (spin singlet) ground state and for the triplet. Again,
and essentially for the same reason, WDFT shows its worth here ---see
Section~\ref{sec:NWO}.

\vspace{6pt}

The customary plan of the paper follows. In Section~2 we briefly
recall for the benefit of the reader our treatment for the singlet
ground state; this helps to introduce the notation. Sections 3 and~4
deal with the general mathematical structure of triplet 1-body Wigner
functions. Section~5 computes the Wigner quasiprobabilities for the
harmonium triplet. Section~6 deals with the corresponding natural
orbitals. In Section~7 the behaviour of the occupation numbers,
obtained numerically, is compared to that of the ground state.
Section~8 continues this comparison in the setting of quantum
information theory. The relative correlation energy for the triplet is
smaller than for the singlet, just as is the purity parameter. The
proportionality between entropy and correlation energy, observed in
the weak correlation limit for the singlet, fails for the triplet
state. Section~9 is the conclusion.

\section{Wigner natural orbitals for the harmonium ground state}
\label{sec:harmonium}

Given any interference operator $\ketbra{\Psi}{\Phi}$ acting on the
Hilbert space of a two-electron system, we denote
\begin{align}
& P_{2\,\Psi\Phi}(\rr_1,\rr_2; \pp_1,\pp_2;
\vs_1,\vs_2;\vs_{1'},\vs_{2'})
\label{eq:basica} 
\\
&\quad := \int \Psi(\rr_1-\zz_1, \rr_2-\zz_2;\vs_1,\vs_2) \,
\Phi^*(\rr_1+\zz_1, \rr_2+\zz_2;\vs_{1'},\vs_{2'}) \,
e^{2i(\pp_1\cdot\zz_1 + \pp_2\cdot\zz_2)} \, d\zz_1\,d\zz_2.
\notag
\end{align}
These are $4 \x 4$ matrices on spin space. When $\Psi = \Phi$ we speak
of Wigner quasiprobabilities, which are always real, and we write
$d_2$ for~$P_2$. The extension of this definition to mixed states is
immediate. The corresponding reduced $1$-body functions are found by
\[
P_{1\,\Psi\Phi}(\rr_1;\pp_1;\vs_1;\vs_{1'}) = 2 \int
P_{2\,\Psi\Phi}(\rr_1,\rr_2;\pp_1,\pp_2; \vs_1,\vs_2;\vs_{1'},\vs_2)
\,d\rr_2 \,d\pp_2 \,d\vs_2.
\]
These are $2 \x 2$ matrices on spin space. When $\Psi = \Phi$ we write
$d_1$ for~$P_1$. The associated spinless quantities
$d_2(\rr_1,\rr_2; \pp_1,\pp_2)$ and $d_1(\rr;\pp)$ are obtained by
tracing on the spin variables. The marginals of $d_2$ give the pairs
densities $\rho_2(\rr_1,\rr_2)$, $\pi_2(\pp_1,\pp_2)$. The marginals
of $d_1$ give the electronic density, namely
$\rho(\rr_1) = \int d_1(\rr_1,\pp_1) \,d\pp_1$, and the momentum
density $\pi(\pp_1) = \int d_1(\rr_1,\pp_1) \,d\rr_1$.

It should be obvious how to extend the definitions to $N$-electron
systems and their reduced quantities; the combinatorial factor for
$d_N \mapsto d_n$ is~$\binom{N}{n}$.

Putting together \eqref{eq:margaritas-vestras} and
\eqref{eq:ad-porcos} with~\eqref{eq:basica}, one arrives~\cite{Pluto}
at:
\begin{align}
d_2(\rr_1,\rr_2;\pp_1,\pp_2;\vs_1,\vs_2;\vs_{1'},\vs_{2'})
&= (\text{spin factor}) \x \sum_{ij} \frac{c_i\,c_j}2 \,
\chi_{ij}(\rr_1;\pp_1) \chi_{ij}(\rr_2;\pp_2),
\label{eq:rho} 
\\[-\jot]
\text{and}\quad  d_1(\rr_1;\pp_1;\vs_1,\vs_{1'})
&= 2 \int d_2(\rr_1,\rr_2;\pp_1,\pp_2; \vs_1,\vs_2;\vs_{1'},\vs_2)
\,d\vs_2 \,d\rr_2 \,d\pp_2
\notag \\
&= \bigl( \up_1\up_{1'} + \dn_1\dn_{1'} \bigr)
\sum_i n_i\, \chi_i(\rr_1;\pp_1).
\notag
\end{align}
Here $n_i$ are the occupation numbers with $\sum_i n_i = 1$, the
$\chi_{ij}$ the natural Wigner interferences and $\chi_i := \chi_{ii}$
denote the natural Wigner orbitals; the spin factor is that
of~\eqref{eq:margaritas-vestras}. Evidently
$\bigl( \up_1\up_{1'} + \dn_1\dn_{1'} \bigr)$ is a rotational scalar.
We replace it by~$2$ in what follows.

The relation $c_i = \pm \sqrt{n_i}$ holds. In principle there still
remains the problem of determining the signs of the infinite set of
square roots, to find the ground state. To recover $d_2$ from~$d_1$ is
no mean feat, since it involves going from a statistical mixture to a
pure state ---see below.

\smallskip

Bringing in extracule and intracule coordinates, respectively given by
\begin{align*}
\RR &= \frac{1}{\sqrt{2}}(\rr_1 + \rr_2), \qquad \rr =
\frac{1}{\sqrt{2}}(\rr_1 - \rr_2),
\\
\PP &= \frac{1}{\sqrt{2}}(\pp_1 + \pp_2), \qquad \pp =
\frac{1}{\sqrt{2}}(\pp_1 - \pp_2),
\end{align*}
the harmonium Hamiltonian is rewritten:
\[
H = H_R + H_r := \frac{P^2}{2} + \frac{\om^2 R^2}{2} + \frac{p^2}{2}
+ \frac{\mu^2 r^2}{2}.
\]
We have introduced the frequencies $\om := \sqrt k$ and
$\mu := \sqrt{k - \dl}$. Assume $\dl < k$, so both ``electrons''
remain in the potential well. For the harmonium ground state the
(spinless) Wigner 2-body quasiprobability is readily found
\cite{Dahl09}:
\begin{align}
d_2(\rr_1,\rr_2;\pp_1,\pp_2) &= \frac{1}{\pi^6}
\exp\biggl(-\frac{2H_R}\om \biggr) \exp\biggl( -\frac{2H_r}\mu
\biggr).
\label{eq:west-ham} 
\end{align}
The reduced 1-body phase space quasiprobability for the ground state
is thus obtained:
\[
d_1(\rr_1;\pp_1) = \frac{2}{\pi^3} \biggl(
\frac{4\om\mu}{(\om + \mu)^2} \biggr)^{3/2}
e^{-2r_1^2\om\mu/(\om + \mu)} e^{-2p_1^2/(\om + \mu)}.
\]
For its natural orbital expansion, with $i$ integer $\geq 0$ and $L_i$
the corresponding Laguerre polynomial, one finds~\cite{Pluto}
\begin{align}
c_i^2
&= n_i = \frac{4\sqrt{\om\mu}}{\bigl(\sqrt\om + \sqrt\mu\,\bigr)^2}
\biggl(\frac{\sqrt\om - \sqrt\mu}{\sqrt\om + \sqrt\mu} \biggr)^{2i}
=: (1 - t^2)\,t^{2i} \,;
\label{eq:harm-param} 
\\
f_i(\rr_1;\pp_1) &= f_i(x_1;p_{1x}) f_i(y_1;p_{1y}) f_i(z_1;p_{1z}),
\word{where}
\notag \\
f_i(x;p_x) &= \frac{1}{\pi}\, (-1)^i
L_i\bigl( 2\sqrt{\om\mu}\,x^2 + 2p_x^2/\sqrt{\om\mu} \bigr) 
e^{-\sqrt{\om\mu}\,x^2 - p_x^2/\sqrt{\om\mu}}.
\notag
\end{align}
The functions $f_i$ determine up to a phase the interferences:
for $j \geq k$,
\begin{align*}
f_{jk}(x,p_x) &= \frac{1}{\pi}\, (-1)^k \frac{\sqrt{k!}}{\sqrt{j!}}\,
\bigl( 2\sqrt{\om\mu}\,x^2 + 2p_x^2/\sqrt{\om\mu} \bigr)^{(j-k)/2}
\\
&\quad \x e^{-i(j - k)\vth}
L_k^{j-k} \bigl( 2\sqrt{\om\mu}\,x^2 + 2p_x^2/\sqrt{\om\mu} \bigr)
e^{-\sqrt{\om\mu}\,x^2 - p_x^2/\sqrt{\om\mu}},
\end{align*}
where
$$
\vth := \arctan\bigl( p_x/\!\sqrt{\om\mu}\,x \bigr).
$$
The $L_k^{j-k}$ are associated Laguerre polynomials. The $f_{kj}$ are
complex conjugates of the~$f_{jk}$. Now, with the \textit{alternating
choice} (unique up to a global sign):
\[
c_i = (-)^i\,\sqrt{n_i} = \sqrt{1 - t^2}\,(-t)^i,
\]
and the above $f_{jk}$, formula \eqref{eq:rho} does reproduce
\eqref{eq:west-ham}. This was originally proved in~\cite{Pluto}, and
verified by minimization in~\cite{Pallene}; we refer the reader to
those papers. Trivially, the same sign rule holds for natural orbitals
of the garden variety~\eqref{eq:margaritas-vestras}.

\section{Generalities on the triplet state}
\label{sec:Triplet}

For a general two-electron system in a triplet spin state the reduced
1-density possesses three different spin factors, say
$$
\up_1\up_{1'} \word{and} 
\half \bigl( \up_1\up_{1'} + \dn_1\dn_{1'} \bigr)  \word{and}
\dn_1\dn_{1'} \,.
$$
While the spatial function for the ground state is symmetric, and
consequently its spin part antisymmetric, for the first excited state
the situation is exactly the opposite: the spatial function is
antisymmetric and its spin part is symmetric. This leads to important
differences between both cases for the natural orbital decomposition.

General triplet states are describable in the form
\cite{LoewdinS56,Davidson76}:
\begin{align*}
\Psi_{t1}(\rr_1, \rr_2; \vs_1,\vs_2)
&= \up_1\up_2 \sum_{ij} \frac{1}{2} c_{ij}
\,[\psi_i(\rr_1) \psi_j(\rr_2) - \psi_j(\rr_1) \psi_i(\rr_2)],
\\
\Psi_{t0}(\rr_1, \rr_2; \vs_1,\vs_2)
&= \frac{1}{\sqrt{2}} \bigl( \up_1\dn_2 + \dn_1\up_2 \bigr)
\sum_{ij} \frac{1}{2} c_{ij}
\,[\psi_i(\rr_1) \psi_j(\rr_2) - \psi_j(\rr_1) \psi_i(\rr_2)],
\\
\Psi_{t,-1}(\rr_1, \rr_2; \vs_1,\vs_2)
&= \dn_1\dn_2 \sum_{ij} \frac{1}{2} c_{ij}
\,[\psi_i(\rr_1) \psi_j(\rr_2) - \psi_j(\rr_1) \psi_i(\rr_2)],
\end{align*}
where $c_{ij} = -c_{ji}$. Here $\{\psi_i\}$ is a complete orthonormal
set. In the absence of magnetic fields, the wave functions can be
taken real. We thus assume that the matrix $C = [c_{ij}]$ is real, as
well as the functions $\psi_i$. Wave function normalization gives rise
to $\Tr(C^t\,C) = \sum_{ij} c_{ij}^2 = 1$.

For the spin part, a less conventional and more cogent description is
found in terms of polarization vectors and the correlation tensor
\cite[App.~F]{Blum12}; however, it is hardly worthwhile to introduce it
here. So we shall be content with presenting the Wigner 2-body
quasiprobabilities for triplet states under the matrix form
\begin{gather*}
P_{2\,\Psi_{t1}\Psi_{t1}} = \up_1\up_2 \up_{1'}\up_{2'} \,d_2
= \begin{pmatrix}
d_2 & 0 & 0 & 0 \\
0 & 0 & 0 & 0 \\
0 & 0 & 0 & 0 \\
0 & 0 & 0 & 0 \end{pmatrix}, \quad
P_{2\,\Psi_{t,-1}\Psi_{t,-1}} = \dn_1\dn_2 \dn_{1'}\dn_{2'} \,d_2
= \begin{pmatrix}
0 & 0 & 0 & 0 \\
0 & 0 & 0 & 0 \\
0 & 0 & 0 & 0 \\
0 & 0 & 0 & d_2 \end{pmatrix},
\\
P_{2\,\Psi_{t0}\Psi_{t0}}
= \frac{1}{2} \bigl( \up_1\dn_2 + \dn_1\up_2\bigr)
\bigl(\up_{1'}\dn_{2'} + \dn_{1'}\up_{2'}\bigr) \,d_2
= \frac{1}{2} \begin{pmatrix}
0 & 0 & 0 & 0 \\
0 & d_2 & d_2 & 0 \\
0 & d_2 & d_2 & 0 \\
0 & 0 & 0 & 0 \end{pmatrix};
\end{gather*}
where $d_2$ is the spinless Wigner 2-body quasiprobability, given by
the expression
\begin{align}
d_2(\rr_1,\rr_2; & \,\pp_1,\pp_2)
\notag \\
&= \frac{1}{4} \sum_{ij,kl} c_{ij}\,c_{kl} \int
[\psi_i(\rr_1 - \zz_1) \psi_j(\rr_2 - \zz_2)
- \psi_j(\rr_1 - \zz_1) \psi_i(\rr_2 - \zz_2)]
\notag \\
&\qquad \x [\psi_k^*(\rr_1 + \zz_1) \psi_l^*(\rr_2 + \zz_2)
- \psi_l^*(\rr_1 + \zz_1) \psi_k^*(\rr_2 + \zz_2)]\,
e^{2i(\pp_1\.\zz_1 + \pp_2\.\zz_2)} \,d\zz_1 \,d\zz_2
\notag \\
&= \frac{1}{4} \sum_{ij,kl} c_{ij}\,c_{kl}
\,[P_{ik}(\rr_1;\pp_1) P_{jl}(\rr_2;\pp_2)
- P_{il}(\rr_1;\pp_1) P_{jk}(\rr_2;\pp_2)
\notag \\[-2\jot]
&\hspace*{7em} - P_{jk}(\rr_1;\pp_1) P_{il}(\rr_2;\pp_2)
+ P_{jl}(\rr_1;\pp_1) P_{ik}(\rr_2;\pp_2)].
\label{eq:rhodos-fs} 
\end{align}
By integrating out one set of coordinates, we obtain the 1-body
quasiprobabilities:
\begin{gather*}
P_{1\,\Psi_{t1}\Psi_{t1}}
= \up\up' \,d_1 = \twobytwo{d_1}{0}{0}{0},  \qquad
P_{1\,\Psi_{t,-1}\Psi_{t,-1}}
= \dn\dn' \,d_1 = \twobytwo{0}{0}{0}{d_1},
\\
P_{1\,\Psi_{t0}\Psi_{t0}}
= \frac{1}{2} \bigl(\up\up' + \dn\dn' \bigr) \,d_1
= \frac{1}{2} \twobytwo{d_1}{0}{0}{d_1}.
\end{gather*}
Here $d_1$ is the spinless 1-body quasidensity corresponding to the
triplet:
\begin{align*}
d_1 (\rr; \pp)
&= 2 \int d_2(\rr,\rr_2; \pp,\pp_2) \,d\rr_2 \,d\pp_2 
\\
&= \frac{1}{2} \sum_{ij,kl} c_{ij}\,c_{kl} \int
[P_{ik}(\rr;\pp) P_{jl}(\rr_2;\pp_2)
- P_{il}(\rr;\pp) P_{jk}(\rr_2;\pp_2)
\\[-2\jot]
&\hspace*{8em} - P_{jk}(\rr;\pp) P_{il}(\rr_2;\pp_2)
+ P_{jl}(\rr;\pp) P_{ik}(\rr_2;\pp_2)] \,d\rr_2 \,d\pp_2
\\
&= \frac{1}{2} \sum_{ij,kl} c_{ij}\,c_{kl}
[P_{ik}(\rr;\pp) \,\dl^j_l - P_{il}(\rr;\pp) \,\dl^j_k
- P_{jk}(\rr;\pp) \,\dl^i_l + P_{jl}(\rr;\pp) \,\dl^i_k] 
\\
&= 2 \sum_{ij,k} c_{ik}\,c_{jk} \, P_{ij}(\rr;\pp)
=  2 \sum_{ij} d_{ij}\, P_{ij}(\rr;\pp),
\end{align*}
where $D = CC^t = -C^2$ is a positive definite matrix.

\section{The Schmidt decomposition of the triplet}
\label{sec:Schmidt}

Let $C$ be any real antisymmetric square matrix. It is well known that
there exists a real orthogonal matrix $Q$ such that $A = Q^tCQ$, with
$A$ a real block-diagonal matrix:
\[
A = \diag[A_0, A_1, \dots],  \qquad  A_0 = 0,  \qquad
A_i = \twobytwo{0}{a_i}{-a_i}{0}.
\]
By convention, here $a_i \geq 0$. Therefore 
\begin{align*}
\sum_{ij,kl} & \,c_{ij}c_{kl} P_{ik}(\rr_1;\pp_1) P_{jl}(\rr_2;\pp_2)
\\ 
&= \sum_{ij,kl,vw}a_v a_w 
\bigl[ q_{i,2v} q_{j,2v+1} - q_{i,2v+1} q_{j,2v} \bigr]
\bigl[ q_{k,2w} q_{l,2w+1} - q_{k,2w+1} q_{l,2w} \bigr]
P_{ik}(\rr_1;\pp_1) P_{jl}(\rr_2;\pp_2)
\\
&= \sum_{ij,kl,vw} a_{v} a_w \bigl[ q_{i,2v} P_{ik}(\rr_1;\pp_1)
q_{k,2w} q_{j,2v+1} P_{jl}(\rr_2;\pp_2) q_{l,2w+1} 
\\[-3\jot]
&\hspace*{8em} - q_{i,2v}  P_{ik}(\rr_1;\pp_1) q_{k,2w+1} q_{j,2v+1} 
P_{jl}(\rr_2;\pp_2) q_{l,2w} 
\\
&\hspace*{8em} - q_{i,2v+1}  P_{ik}(\rr_1;\pp_1) q_{k,2w} q_{j,2v}
P_{jl}(\rr_2;\pp_2) q_{l,2w+1} 
\\
&\hspace*{8em} + q_{i,2v+1}  P_{ik}(\rr_1;\pp_1) q_{k,2w+1} q_{j,2v}
P_{jl}(\rr_2;\pp_2) q_{l,2w}  \bigr].
\end{align*}
Let us now make the definition
$\chi_{rp}(\rr;\pp) := \sum_{mk} q_{mr}\, P_{mk}(\rr;\pp) \,q_{kp}$,
so that
$P_{mk}(\rr;\pp) = \sum_{rp} q_{mr}\, \chi_{rp}(\rr;\pp) \,q_{kp}$.
This is the set of Wigner natural orbitals, and has the following
nice property:
\[
\int \chi_{rp}(\rr;\pp) \,d\rr\,d\pp 
= \int \sum_{mk} q_{mr} P_{mk}(\rr;\pp) q_{kp} \,d\rr \,d\pp 
= \sum_{mk} q_{mr} q_{kp} \,\delta^m_k = \delta^r_p .
\]
Hence,
\begin{align*}
\sum_{ij,kl} & \,c_{ij}c_{kl} P_{ik}(\rr_1;\pp_1) P_{jl}(\rr_2;\pp_2)
\\
&= \sum_{vw} a_v a_w \bigl[ 
\chi_{2v,2w}(\rr_1;\pp_1) \chi_{2v+1,2w+1}(\rr_2;\pp_2)
- \chi_{2v,2w+1}(\rr_1;\pp_1) \chi_{2v+1,2w}(\rr_2;\pp_2)
\\
&\hspace*{5em}
- \chi_{2v+1,2w}(\rr_1;\pp_1) \chi_{2v,2w+1}(\rr_2;\pp_2)
+ \chi_{2v+1,2w+1}(\rr_1;\pp_1) \chi_{2v,2w}(\rr_2;\pp_2) \bigr].
\end{align*}

The other three summands in \eqref{eq:rhodos-fs} yield the same
expression. For instance, the third is
\begin{align*}
- & \sum_{ij,kl} c_{ij}c_{kl} P_{il}(\rr_1;\pp_1) P_{jk}(\rr_2;\pp_2)
\\
&= - \!\sum_{ij,kl,vw} a_v a_w 
\bigl[ q_{i,2v} q_{j,2v+1} - q_{i,2v+1} q_{j,2v} \bigr] 
\bigl[ q_{k,2w} q_{l,2w+1} - q_{k,2w+1} q_{l,2w} \bigr]
P_{il}(\rr_1;\pp_1) P_{jk}(\rr_2;\pp_2)
\\
&= - \!\sum_{ij,kl,vw} a_v a_w \bigl[q_{i,2v} P_{il}(\rr_1;\pp_1)
q_{l,2w+1} q_{j,2v+1} P_{jk}(\rr_2;\pp_2) q_{k,2w} 
\\[-3\jot]
&\hspace*{8em} - q_{i,2v} P_{il}(\rr_1;\pp_1) q_{l,2w} 
 q_{j,2v+1} P_{jk}(\rr_2;\pp_2) q_{k,2w+1} 
\\
&\hspace*{8em} - q_{i,2v+1} P_{il}(\rr_1;\pp_1) q_{l,2w+1} 
q_{j,2v} P_{jk}(\rr_2;\pp_2) q_{k,2w} 
\\
&\hspace*{8em} + q_{i,2v+1} P_{il}(\rr_1;\pp_1) q_{l,2w} 
q_{j,2v} P_{jk}(\rr_2;\pp_2) q_{k,2w+1} \bigr].
\end{align*}
This leads to the same contribution as the first summand. Then use
symmetry under the interchange of the two particles. In summary,
\begin{align}
& d_2(\rr_1,\rr_2; \pp_1,\pp_2)
\notag \\
&\quad = \sum_{vw} a_v a_w
\bigl[ \chi_{2v,2w}(\rr_1;\pp_1) \chi_{2v+1,2w+1}(\rr_2;\pp_2)
- \chi_{2v,2w+1}(\rr_1;\pp_1) \chi_{2v+1,2w}(\rr_2;\pp_2)
\notag \\
&\hspace*{6em}
- \chi_{2v+1,2w}(\rr_1;\pp_1) \chi_{2v,2w+1}(\rr_2;\pp_2)
+ \chi_{2v+1,2w+1}(\rr_1;\pp_1) \chi_{2v,2w}(\rr_2;\pp_2) \bigr].
\label{eq:d2-final-expr} 
\end{align}

The reduced 1-body phase space (spinless) quasidensity for the
triplet is obtained, as before,
\begin{equation}
d_1(\rr;\pp) = 2 \int d_2(\rr,\rr_2;\pp,\pp_2) \,d\rr_2 \,d\pp_2
= 2 \sum_w a_w^2 \,[\chi_{(2w,2w)}(\rr;\pp)
+ \chi_{(2w+1,2w+1)}(\rr;\pp)].
\label{eq:d1-first-state} 
\end{equation}
Notice that in the previous equation each occupation number
$n_i := 2a_i^2$ appears twice. This is a consequence of the Pauli
exclusion principle.

Unlike the singlet case, there is \textit{no sign rule} to be
deciphered here. Instead there are the ambiguities:
\begin{align*}
\chi_{2w,2w} &= \chi'_{2w,2w} \cos^2\th_w
- (\chi'_{2w,2w+1} + \chi'_{2w+1,2w}) \sin\th_w \cos\th_w 
+ \chi'_{2w+1,2w+1} \sin^2\th_w,
\\
\chi_{2w+1,2w+1} &= \chi'_{2w,2w} \sin^2\th_w
+ (\chi'_{2w,2w+1} + \chi'_{2w+1,2w}) \sin\th_w \cos\th_w
+ \chi'_{2w+1,2w+1} \cos^2\th_w.
\end{align*}
They clearly leave the form \eqref{eq:d1-first-state} untouched. We
see here the action of $SO(2)$ on each invariant block. One may choose
the angles as to maximize their overlap with the leading natural
orbitals for the ground state, as done in the seminal paper by
L\"owdin and Shull~\cite{LoewdinS56}. We omit that. Let us define
\[
A_w := \twobytwo{\cos\th_w}{-\sin\th_w}{\sin\th_w}{\cos\th_w}.
\]
The above transformation can be construed as
\begin{align*}
\chi = (A_v \ox A_w)\,\chi' = \begin{pmatrix}
 \cos\th_v \cos\th_w & -\cos\th_v \sin\th_w &
-\sin\th_v \cos\th_w &  \sin\th_v \sin\th_w \\
 \cos\th_v \sin\th_w &  \cos\th_v \cos\th_w &
-\sin\th_v \sin\th_w & -\sin\th_v \cos\th_w \\
 \sin\th_v \cos\th_w & -\sin\th_v \sin\th_w &
 \cos\th_v \cos\th_w & -\cos\th_v \sin\th_w \\
 \sin\th_v \sin\th_w &  \sin\th_v \cos\th_w &
 \cos\th_v \sin\th_w &  \cos\th_v \cos\th_w \end{pmatrix} \chi',
\end{align*}
with
$$
\chi := \begin{pmatrix} \chi_{2v,2w} \\ \chi_{2v,2w+1} \\
\chi_{2v+1,2w} \\ \chi_{2v+1,2w+1} \end{pmatrix},
\quad \text{and similarly for } \chi',
$$
in the case $v = w$.

To similarly examine the symmetry of expression
\eqref{eq:d2-final-expr}, again one does not have to contend with the
whole tensor product matrix, since most contributions vanish. As
regards the sum in~\eqref{eq:d2-final-expr}, one can write in
compressed form:
\begin{align*}
\chi \chi = \begin{pmatrix}
 \cos^2\th_v \cos^2\th_w & -\cos^2\th_v \sin^2\th_w &
-\sin^2\th_v \cos^2\th_w &  \sin^2\th_v \sin^2\th_w \\
-\cos^2\th_v \sin^2\th_w &  \cos^2\th_v \cos^2\th_w &
 \sin^2\th_v \sin^2\th_w & -\sin^2\th_v \cos^2\th_w \\
-\sin^2\th_v \cos^2\th_w &  \sin^2\th_v \sin^2\th_w &
 \cos^2\th_v \cos^2\th_w & -\cos^2\th_v \sin^2\th_w \\
 \sin^2\th_v \sin^2\th_w & -\sin^2\th_v \cos^2\th_w &
-\cos^2\th_v \sin^2\th_w &  \cos^2\th_v \cos^2\th_w \end{pmatrix}
\chi' \chi',
\end{align*}
with
$$
\chi \chi := \begin{pmatrix} 
\chi_{2v,2w}(\rr_1;\pp_1) \,\chi_{2v+1,2w+1}(\rr_2;\pp_2) \\ 
\chi_{2v,2w+1}(\rr_1;\pp_1) \,\chi_{2v+1,2w}(\rr_2;\pp_2) \\ 
\chi_{2v+1,2w}(\rr_1;\pp_1) \,\chi_{2v,2w+1}(\rr_2;\pp_2) \\
\chi_{2v+1,2w+1}(\rr_1;\pp_1) \,\chi_{2v,2w}(\rr_2;\pp_2)
\end{pmatrix}; \word{and similarly for} \chi'\chi'.
$$
One verifies that \eqref{eq:d2-final-expr} is invariant under this set
of transformations.

\section{Lowest triplet state of harmonium}
\label{sec:TripletMore}

The energy spectrum for harmonium is obviously
$(\N + \frac{3}{2})\om + (\N + \frac{3}{2})\mu$. Since $\mu < \om$,
the energy of the first excited states is $E_\fs =  (3\om + 5\mu)/2$.
For our present purposes, it is enough to choose an intracule
excitation state along the $x$-axis (say). The corresponding
2-quasidensity is given by:
\begin{align}
\frac{2}{\pi^6} \exp\biggl( -\frac{2H_R}{\om} \biggr)
\exp\biggl( -\frac{2H_r}{\mu} \biggr)
\biggl( \frac{(p_{1x} - p_{2x})^2 + \mu^2(x_1^2 - x_2^2)^2}{\mu}
- \frac{1}{2} \biggr).
\label{eq:wigner-fs} 
\end{align}

Henceforth we work in the chosen nontrivial mode, since the problem
factorizes completely. By integrating one set of variables, the
reduced one-body spinless quasidensity is obtained, after some work:
\begin{equation}
d_1(r;p) = 2 \int d_2(r,r_2;p,p_2) \,dr_2 \,dp_2 
= \frac{2}{\pi} \biggl( \frac{2\sqrt{\om\mu}}{\om + \mu} \biggr)^3
e^{-\frac{2\om\mu}{\om+\mu} r^2 - \frac{2}{\om+\mu} p^2}
\biggl( \om r^2 + \frac{1}{\om} p^2 \biggr).
\label{eq:first-state} 
\end{equation}

The marginals of $d_1$ give the electronic density and momentum
density:
\begin{align*}
\rho(r) = \int d_1(r;p) \,dp
&= \frac{2}{\pi} \biggl( \frac{2\sqrt{\om\mu}}{\om + \mu} \biggr)^3
e^{-\frac{2\om\mu}{\om+\mu} r^2} \int e^{-\frac{2}{\om+\mu} p^2}
\biggl(\om r^2 + \frac{1}{\om} p^2 \biggr) \,dp
\\
&= \frac{2}{\pi} \biggl( \frac{2\sqrt{\om\mu}}{\om + \mu} \biggr)^3
e^{-\frac{2\om\mu}{\om+\mu} r^2}
\biggl( \frac{\pi(\om + \mu)}{2} \biggr)^{1/2}
\biggl( \om r^2 + \frac{\om+\mu}{4\om} \biggr),
\\
\pi(p) = \int d_1(r;p) \,dr
&= \frac{2}{\pi} \biggl( \frac{2\sqrt{\om\mu}}{\om + \mu} \biggr)^3
e^{-\frac{2}{\om+\mu} p^2} \int e^{-\frac{2\om\mu}{\om+\mu}r^2}
\biggl( \om r^2 + \frac{1}{\om} p^2 \biggr) \,dr
\\
&= \frac{2}{\pi} \biggl( \frac{2\sqrt{\om\mu}}{\om + \mu} \biggr)^3
e^{-\frac{2}{\om+\mu} p^2}
\biggl( \frac{\pi(\om + \mu)}{2\om\mu} \biggr)^{1/2}
\biggl( \frac{\om+\mu}{4\mu} + \frac{1}{\om} p^2 \biggr).
\end{align*}
Finally, as expected, we get
\begin{equation*}
\int \pi(p) \,dp = \int \rho(r) \,dr
= \frac{2}{\pi} \biggl( \frac{2\sqrt{\om\mu}}{\om + \mu} \biggr)^3
\biggl( \frac{\pi(\om + \mu)}{2} \biggr)^{1/2}
\biggl( \frac{\pi(\om + \mu)}{2\om\mu} \biggr)^{1/2}
\biggl( \frac{\om + \mu}{4\mu} + \frac{\om + \mu}{4 \om} \biggr) = 2.
\end{equation*}

{}From the viewpoint of WDFT, the most interesting part of the energy
corresponds to the interelectronic repulsion of this first excited
state $E_{2\fs}$. The 1-body Hamiltonian is given by
$h(r,p) = p^2/2 + \om^2 r^2/2$. It is a simple exercise to obtain the
1-body energy $E_{1\fs}$ by integrating
expression~\eqref{eq:first-state} with this observable:
\[
E_{1\fs} = \frac{\om}{2} + \frac{3(\mu^2 + \om^2)}{4\mu}.
\]
The interelectronic potential in~\eqref{eq:Mosh-atom} is
$(\mu^2 - \om^2) r_{12}^2/4$, so to obtain the repulsion energy
$E_{2\fs}$, one has just to integrate expression~\eqref{eq:wigner-fs}
with this observable:
\begin{align*}
E_{2\fs} &= \int \frac{2}{\pi^2} \exp\biggl( -\frac{2H_R}{\om} \biggr)
\exp\biggl( -\frac{2H_r}{\mu} \biggr)
\biggl[ \frac{2H_r}{\mu} - \frac{1}{2}\biggr]
\frac{\mu^2 - \om^2}{4}\, r_{12}^2 \,dR \,dr \,dP \,dp
\\
&= \frac{1}{\pi} (\mu^2 - \om^2) \int
\exp\biggl( -\frac{2H_r}{\mu} \biggr)
\biggl[ \frac{r^2 p^2}{\mu} + \mu r^4 - \frac{r^2}{2} \biggr] \,dr\,dp
= \frac{3}{4}\, \frac{\mu^2 - \om^2}{\mu},
\end{align*}
which is $3$~times the interelectronic repulsion energy for the
corresponding mode of the singlet~\cite{Pluto}. This is not
surprising, since, in the triplet configuration the electrons tend to
be mutually farther apart than in the singlet.%
\footnote{Interestingly, \eqref{eq:first-state} is a non-Gaussian
Wigner function taking only positive values. This prompts two remarks.
First, in consonance with common wisdom \cite{KenfackZ04,DahlMWS06},
it is confirmed that as of itself $d_1$ is a nearly classical state.
Second, there are mathematical recipes that produce such
positive-valued Wigner functions representing mixed
states~\cite{Titania}. It would be good to know whether or not
\eqref{eq:first-state} can be obtained as such an output.}

\section{Spectral analysis of the 1-body triplet state}
\label{sec:NWO}

In order to determine the occupation numbers of this system, first we
have to find the good coordinates. Let us perform the transformation
$$
(Q,P) := \bigl( (\om\mu)^{1/4}r, (\om\mu)^{-1/4}p \bigr);
\word{or, in shorthand,}  U = Su,
$$
where $S$ is symplectic and $u = (r,p)$. We may also write
$\vth := \arctan(P/Q)$, so that
$$
P = U \sin\vth  \word{and}  Q = U \cos\vth.
$$
Recalling $2\sqrt{\om\mu}/(\om + \mu) = (1 - t^2)/(1 + t^2)$
from~\eqref{eq:harm-param}, the 1-quasidensity \eqref{eq:first-state}
takes the simple form:
\begin{align*}
d_1(U, \vth) := d_1(u(U,\vth))
&= \frac{2(1 - t^2)^3}{\pi(1 + t^2)^3}\, e^{-(1 - t^2)U^2/(1 + t^2)}
U^2 \biggl( \frac{1 + t}{1 - t}\cos^2\vth
+ \frac{1 - t}{1 + t}\sin^2\vth \biggr)
\\
&= \frac{2(1 - t^2)^3}{\pi(1 + t^2)^3}\, e^{-(1 - t^2)U^2/(1 + t^2)}
U^2 \biggl( \frac{1 + t^2}{1 - t^2} + \frac{2t}{1 - t^2} \cos 2\vth
\biggr).
\end{align*}

The one-body quasidensity may be expanded as follows:
$$
d_1(U,\vth) = \sum_{rs} f_{rs}(U,\vth)\,d_{rs}   \word{where}
d_{rs} = 2\pi \int d_1(U,\vth) f^*_{rs}(U,\vth) U \,dU \,d\vth.
$$
Then, for $r \geq s$,
\begin{align*}
2\pi \int & f^*_{rs}(U,\vth)\, d_1(U, \vth)\, U \,dU \,d\vth
\\
&= \frac{4(1 - t^2)^3}{\pi(1+t^2)^3}\, (-1)^s
\frac{\sqrt{s!}}{\sqrt{r!}} \int_0^\infty e^{-(1 - t^2)U^2/(1 + t^2)}
e^{-U^2} (2U^2)^{(r-s)/2} L_s^{r-s}(2U^2)\, U^3 \,dU
\\
&\qquad \x \int_{-\pi}^\pi e^{i(r-s)\vth} \biggl[ 
\frac{1 + t^2}{1 - t^2} + \frac{2t}{1 - t^2}\cos 2\vth \biggr] \,d\vth
\\
&= \frac{4(1 - t^2)^3}{\pi(1+t^2)^3}\, (-1)^s
\frac{\sqrt{s!}}{\sqrt{r!}} \int_0^\infty e^{-(1 - t^2)U^2/(1 + t^2)}
e^{-U^2} (2U^2)^{(r-s)/2} L_s^{r-s}(2U^2)\, U^3 \,dU
\\
&\qquad \x \pi \biggl[ \frac{2(1 + t^2)}{1 - t^2} \delta_r^s
+ \frac{2t}{1 - t^2} (\delta^{s+2}_r + \delta^{s-2}_r) \biggr],
\end{align*}
so that
$$
d_1(U,\vth) = \sum_s d_{ss}(t) f_{ss}(U,\vth)
+ d_{s+2,s}(t) f_{s+2,s}(U,\vth) + d_{s,s+2}(t)f_{s,s+2}(U,\vth),
$$
where actually $d_{s+2,s} = d_{s,s+2}$.

Using the standard Mellin transform \cite{PrudnikovBM83,AndrewsAR99}:
$$
\int_0^\infty x^{\al-1}\, e^{-px} L_n^\la(cx) \,dx
= \frac{\Ga(\al)}{p^\al} P_n^{(\la, \al - \la - n - 1)}
\Bigl( 1 - \frac{2c}{p} \Bigr)
= \frac{\Ga(\al)}{p^\al} \frac{(\la + 1)_n}{n!}\,
\dosFuno{-n}{\al}{\la + 1}{\frac{c}{p}},
$$
we obtain by fairly easy manipulations,
\begin{align*}
d_{ss}(t) &= (1 - t^2)^2\bigl( s\,t^{2s-2} + (1 + s)\,t^{2s} \bigr);
\\
d_{s,s+2}(t) &= (1 - t^2)^2 \sqrt{(s+1)(s+2)}\, t^{2s+1}.
\end{align*}
This means that, to find the occupation numbers, one has to
diagonalize a symmetric pentadiagonal matrix:
\begin{equation}
D = (1-t^2)^2 \, \begin{pmatrix}
1             & 0            &  \al_0 t         & 0
& 0           & 0            & \cdots \\
0             & 1 + 2 t^2    &  0               & \al_1 t^3
& 0           & 0            & \cdots \\
\al_0 t       & 0            & 2 t^2 + 3 t^4    & 0
& \al_2 t^5   & 0            & \cdots \\
0             & \al_1 t^3    & 0                & 3 t^4 + 4 t^6
& 0           & \al_3 t^7    & \cdots \\
0             & 0            & \al_2 t^5        & 0
& 4 t^6 + 5 t^8 & 0       & \cdots \\
0             & 0            & 0                & \al_3 t^7
& 0           & 5 t^8 + 6 t^{10}  & \cdots \\
\vdots        & \vdots       & \vdots           & \vdots
& \vdots      & \vdots       & \ddots
\end{pmatrix},
\label{eq:pentamatrix} 
\end{equation}
where $\al_s := \sqrt{(s+1)(s+2)}\,$.

It is readily checked that the trace of this matrix is~$2$, as it
should be. Its eigenspaces split into two parts:
$\ell_2 = V_1 \oplus V_2$, where
$V_1 = \set{\xx : \text{ all } x_{2n} = 0}$ and
$V_2 = \set{\xx : \text{ all } x_{2n+1} = 0}$. They correspond
respectively to the matrices
\begin{align*}
D_\even = (1-t^2)^2 \, \begin{pmatrix}
1         & \al_0 t        & 0              & 0
& 0       & \cdots \\
\al_0 t   & 2 t^2 + 3 t^4  & \al_2 t^5    & 0
& 0       & \cdots \\
0         & \al_2 t^5      & 4 t^6 + 5 t^8  & \al_4 t^9
& 0       & \cdots \\
0         & 0              & \al_4 t^9      & 6 t^{10} + 7 t^{12}
& \al_6 t^{13}  & \cdots \\
0         & 0              & 0              & \al_6 t^{13}
& 8 t^{14} + 9 t^{16} & \cdots \\
\vdots    & \vdots         & \vdots  & \vdots  & \vdots  & \ddots
\end{pmatrix}
\end{align*}
and
\begin{align*}
D_\odd = (1-t^2)^2 \, \begin{pmatrix}
1 + 2 t^2     & \al_1 t^3      & 0                & 0
& 0           & \cdots \\
\al_1 t^3     & 3 t^4 + 4 t^6  & \al_3 t^7        & 0
& 0           & \cdots \\
0             & \al_3 t^7      & 5 t^8 + 6 t^{10} & \al_5 t^{11}
& 0           & \cdots \\
0             & 0                & \al_5 t^{11}
& 7 t^{12} + 8 t^{14} & \al_7 t^{15} & \cdots \\
0             & 0                & 0                & \al_7 t^{15}
& 9 t^{16} + 10 t^{18} & \cdots \\
\vdots        & \vdots    & \vdots    & \vdots   & \vdots   & \ddots
\end{pmatrix}.
\end{align*}
It is easily checked that these matrices have the same set of
eigenvalues, as they should, since the occupation numbers must appear
twice.

\renewcommand{\arraycolsep}{3pt} 

As was shown in Section~\ref{sec:Triplet}, there is a skewsymmetric
matrix $C$ such that $D = C^t C$. This matrix is tridiagonal, and is
the sum of two skew-symmetric matrices whose diagonalization is
trivial:
\begin{align*}
C &= (1-t^2) \, \begin{pmatrix}
0  & -1  & 0           & 0              & 0           & \cdots \\
1  &  0  & 0           & 0              & 0           & \cdots \\
0  &  0  & 0           & -\sqrt 3 t^2   & 0           & \cdots \\
0  &  0  & \sqrt 3 t^2 & 0              & 0           & \cdots \\
0  &  0  & 0           & 0              & 0           & \cdots \\
\vdots        & \vdots    & \vdots    & \vdots   & \vdots    & \ddots
\end{pmatrix}
+
(1-t^2) \, \begin{pmatrix}
0  & 0          & 0           & 0             & 0           & \cdots \\
0  & 0          & \sqrt 2 t   & 0             & 0           & \cdots \\
0  & -\sqrt 2 t & 0           & 0             & 0           & \cdots \\
0  & 0          & 0           & 0             & \sqrt 4 t^3 & \cdots \\
0  & 0          & 0           & -\sqrt 4 t^3  & 0           & \cdots \\
\vdots        & \vdots    & \vdots    & \vdots   & \vdots    & \ddots
\end{pmatrix}
\\
&=: A + B.
\end{align*}
Also, $D$ is the sum of two Hermitian matrices, namely $A^tA + B^tB$,
which is diagonal, and $A^tB + B^tA$.

\renewcommand{\arraycolsep}{5pt} 

\medskip

One is reminded here of the Weyl problem: given two $n \x n$ Hermitian
matrices $A$, $B$ whose spectra are known, what could the spectrum of
their sum $C := A + B$ be? Some facts are clear: with an obvious
notation for the eigenvalues, these must satisfy
$$
c_1 +\cdots+ c_n = a_1 +\cdots+ a_n + b_1 +\cdots+ b_n;  \qquad
c_1 \leq a_1 + b_1;
$$
less clear, but also true, are
$$
c_2 \leq a_1 + b_2;  \quad  c_2 \leq a_2 + b_1;
$$
and so on. The conditions written above are already optimal for $n =
2$. The necessary constraints are all linear homogeneous inequalities,
bounding convex polyhedra. Horn made a conjecture for the general form
of such inequalities, which was eventually proved~\cite{KnutsonT01}.

The pure-state $N$-representability problem in quantum chemistry (or
``quantum marginal problem'', in the jargon of information theory)
should be considered as solved, after the work by Klyachko
\cite{Klyachko06,Klyachko09}. It is of the same type and answered by similar
inequalities. Both questions reduce to finding moment polyhedra for
coadjoint orbits of unitary groups (associated to  pertinent
Hilbert spaces), which are computed by Duistermaat--Heckman measures
\cite{DH82}. A very readable and up-to-date account of all this is
\cite{ChristandlBKW11}. The Hilbert spaces considered are
finite-dimensional. However, the results are valid for finite-rank
approximations in the chemical context, and the patterns of the
inequalities extend in a rather obvious way. Thus it is scarcely
surprising that the Weyl problem surfaces in this simple instance. We
leave for the future consideration of the moment polytopes for the
occupation numbers,%
\footnote{The number of their extremal edges grows very quickly with
$N$ and the rank; this makes for precision, but also for strenuous
work.}
and choose in this paper a direct approach to the eigenpair problem,
completed by numerical analysis.

\medskip

The matrices $D_\even$ and $D_\odd$ are tridiagonal symmetric real
matrices. The general eigenvalue problem for a matrix $T$ of this kind
reduces to solving the following set of recurrence equations:
\[
\begin{pmatrix}
d_0 & t_1 & 0   & 0   & \cdots \\
t_1 & d_1 & t_2 & 0   & \cdots \\
0   & t_2 & d_2 & t_3 & \cdots \\
0   & 0   & t_3 & d_3 & \cdots \\
0   & 0   & 0   & t_4 & \cdots \\
\vdots & \vdots & \vdots & \vdots & \ddots  \end{pmatrix} 
\begin{pmatrix}
\phi_0(n_r) \\  \phi_1(n_r) \\  \phi_2(n_r) \\
\phi_3(n_r) \\  \phi_4(n_r) \\  \vdots  \end{pmatrix}
= n_r \begin{pmatrix}
\phi_0(n_r) \\  \phi_1(n_r) \\  \phi_2(n_r) \\
\phi_3(n_r) \\  \phi_4(n_r) \\  \vdots  \end{pmatrix},
\]
where $n_r$ is an eigenvalue. The general solution is completely given
in terms of the occupation numbers, by the following
formula~\cite[Sect.~5.48]{Wilkinson65}:
$$
\phi_m(\la)
= \frac{\phi_0(\la)}{t_1t_2\dots t_m}\, \det[\la I - T]_{mm},
\word{for each} m \geq 1,
$$
where $[\la I - T]_{mm}$ is the upper left $m \x m$ submatrix of
$(\la I - T)$, and $\phi_0(\la) \neq 0$ is chosen so as to normalize
the eigenvector.

This result implies that $T = Q D Q^t$, where $d_{ij} = n_i \dl^i_j$
is the diagonal matrix whose entries are the eigenvalues and 
$q_{ij} = \phi_i(n_j)$. Since $Q Q^t = Q^t Q = 1$, the following
orthogonality relations hold:
\begin{gather*}
\sum_{r=0}^\infty \phi_m(n_r) \phi_l(n_r) = \delta^m_l, \qquad
\sum_{m=0}^\infty \phi_m(n_r) \phi_m(n_s) = \delta^r_s,
\\
\sum_{r=0}^\infty n_r\, \phi_m(n_r) \phi_l(n_r)
= d_m \delta^m_l + t_m \delta^{m-1}_l.
\end{gather*}

In summary, for $d_1$ we obtain
\begin{align*}
d_1(\.) &= \sum_r n_r \biggl[
\sum_{i=0}^\infty f_{2i,2i}(\.) \,\phi_{\even,i}^2(n_r)
+ \sum_{i=0}^\infty (f_{2i,2i+2} + f_{2i+2,2i})(\.)
\,\phi_{\even,i}(n_r) \phi_{\even,i+1}(n_r)
\\
&\qquad + \sum_{i=0}^\infty f_{2i+1,2i+1}(\.) \,\phi_{\odd,i}^2(n_r)
+ \sum_{i=0}^\infty (f_{2i+1,2i+3} + f_{2i+3,2i+1})(\.)
\,\phi_{\odd,i}(n_r) \phi_{\odd,i+1}(n_r) \biggr].
\end{align*}
Here $n_r$ depends solely on the parameter~$t$
of~\eqref{eq:harm-param}.

\section{Numerical analysis of the occupation numbers}
\label{sec:NWON}

\begin{figure}[ht] 
\centering
\includegraphics[width=9cm]{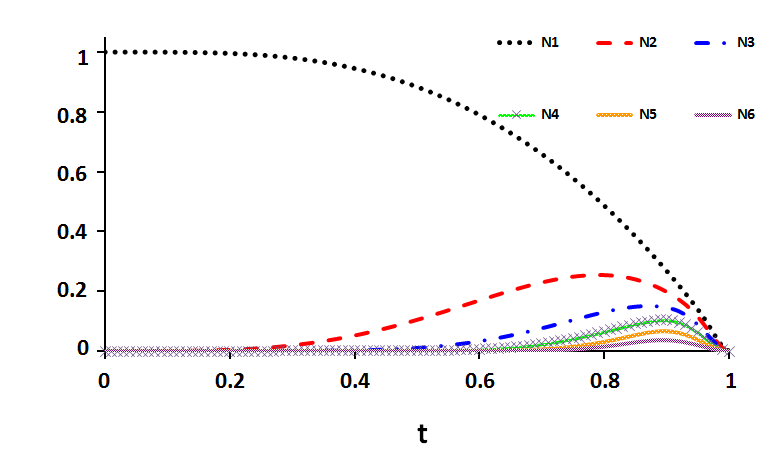}
\caption{First six eigenvalues of the matrix $D_\even$.}
\label{graf:eigenvalues}
\end{figure}

As advertised, to find the $n_r$ we fall back on numerical
computation. Figure~\ref{graf:eigenvalues} shows the behavior of the
rank-eight approximation of the eigenvalues, as $t$ is varied. Note
that the first eigenvalue is very close to~$1$ in the neighborhood of
$t = 0$, while the others are very small. As the value of~$t$ rises,
the first eigenvalue begins to decrease and the others rise for a
while. In the neighborhood of $t = 1$ all eigenvalues approach zero.

Note that $t$ is a very nonlinear parameter: although $t \sim \dl/8k$ 
for small~$\dl$, the value $t = 1/2$ means $\mu/\om = 1/9$ or
$\dl/k = 80/81$. This shows that, unless $\dl$ is pretty close to the
dissociation value, the harmonium triplet is not badly described by a
Hartree--Fock state. Whenever $t \lesssim 0.6$, that is,
$\dl/k \lesssim 255/256$, the first two occupation numbers contain
almost all the physical information for the system.

Also, one we can show that whenever $t \lesssim 0.5$, a good
approximation to the five first occupation numbers is
\[
\la_1 \approx 1 - 3 t^4 + 8 t^6, \quad
\la_2 \approx 3 t^4 - 8 t^6, \quad
\la_3 \approx 5 t^8, \quad
\la_4 \approx 7 t^{12} \word{and}
\la_5 \approx 9 t^{16}.
\]

\begin{figure}[ht] 
\centering
\includegraphics[width=9cm]{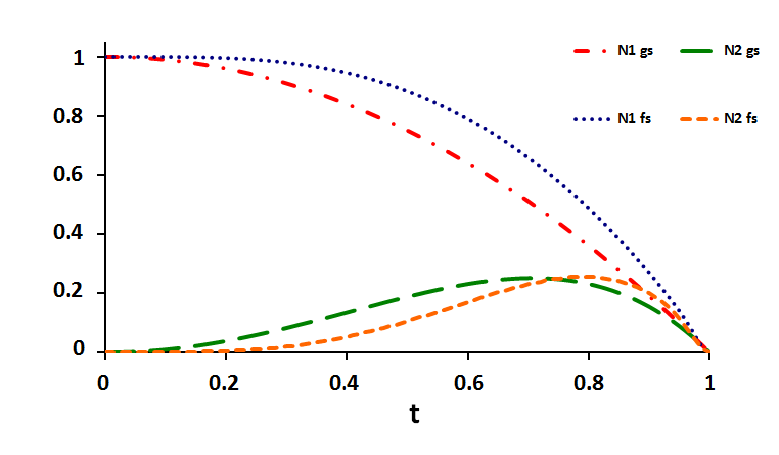}
\caption{First and second occupation numbers of the ground state
and of the first excited state.}
\label{graf:TvS}
\end{figure}

Figure~\ref{graf:TvS} compares the behavior of the first two
eigenvalues for the singlet and triplet states of harmonium. In this
sense, the Hartree--Fock approximation works better in the triplet
case than for the singlet. Around $t = 0.4$ the second approximated
occupation number for the latter is above~$0.13$, and for the former
is below~$0.052$. The same behaviour was also observed in the toy
model studied in~\cite{HelbigTR10}. This does not mean, however, that
correlation is always weaker in the triplet state ---see the next
section.

\section{Spatial entropy and correlation energies}
\label{sec:the-entropy}

We move towards the comparison of the triplet system with the singlet
system in regard to disorder (suppressing the spin variables). To
measure this, a useful quantity is the linear entropy $s$ associated
to the 1-body function:
\[
s = 1 - \Pi(d_1),
\]
where $\Pi(d_1)$ is the \textit{purity} of the system ---see below.%
\footnote{Truth to be told, the notion of entropy native to the Wigner
quasiprobability approach is the one discussed in~\cite{Lieb90}. We
put aside the question of its eventual usefulness here.}
Mathematically, the quantity $s$ is a lower bound for the Jaynes
entropy, which has been used to quantify the entanglement between one
particle and the other $N - 1$ particles of the
system~\cite{HelbigTR10}, and proposed as a handle on the correlation
energies~\cite{SmithSS02}. In this paper the singlet has been modelled
in such a way that, for each one-dimensional mode:
\[
\Pi_{\gs,1}(d_1) = \int d_1^2(r;p)\,dr\,dp = \sum_i n_i^2.
\]
Instead, for the triplet one should take for the excited mode:
\[
\Pi_{\fs,x}(d_1) = \frac{1}{2} \int d_1^2(r_x;p_x) \,dr_x \,dp_x
= \sum_i n_i^2.
\]
This second definition is natural in that correlations due solely to
the antisymmetric character of the wave function \textit{do not
contribute} to the entanglement of the system
\cite{YanyezPD10,NaudtsV07,BalachandranGQRL12}. This ensures that the
entropy for a 1-body function of the Hartree--Fock type is zero.

In the singlet case, the occupation numbers are equal to
$(1 - t^2)\,t^{2i}$. Thus, the purity of this system is easily
computable, to wit, $\Pi_{\gs,1}(d_1) = (1 - t^2)/(1 + t^2)$ for each
mode. This quantity coincides with the quotient of the geometric and
arithmetic means of the frequencies, that is,
$\Pi_{\gs,1} = 2\sqrt{\om\mu}/(\om + \mu)$. For $n$~modes one just
takes the $n$th power~\cite{PipekN09}. Moreover, for small values of
the coupling~$\dl$, we obtain
\begin{equation}
s_{\gs,1} \sim \frac{1}{32}\,\frac{\dl^2}{\om^4},
\label{eq:capeat-qui-potest} 
\end{equation}
which for this approximation is exactly the absolute value of the
(dimensionless) correlation energy~\cite{Hermione}. This appears to
vindicate the contention of~\cite{SmithSS02}. (Actually, for the
singlet it is not difficult to compute the Jaynes entropy, given by
\[
-\sum_i n_i \log n_i = -\log(1 - t^2) - \frac{t^2\log t^2}{1 - t^2}\,.
\]
This was done by Srednicki~\cite{Srednicki93} some time ago.)

\medskip

For the triplet state, we have to compute $\Tr(d_1^2)$ for the matrix 
given in~\eqref{eq:pentamatrix}. Since
\[
d_1^2 = (1-t^2)^4 \, \begin{pmatrix}
1 + \al^2_0 t^2 & 0 & \al_0 t (1 + 2 t^2 + 3 t^4) & \cdots \\
0 & (1 + 2 t^2)^2 + \al^2_1 t^6 & 0 & \cdots \\
\al_0 t (1 + 2 t^2 + 3 t^4)  & 0
& \al^2_0 t^2 + (2 t^2 + 3 t^4)^2 + \al^2_2 t^{10} & \cdots \\
\vdots & \vdots & \vdots & \ddots \end{pmatrix},
\]
we get
\begin{align*}
\Tr(d_1^2) &= (1 - t^2)^4 \biggl[ 
4 \sum_{i=0}^\infty \al_i^2\, t^{2(2i+1)}
+ 2 \sum_{i=1}^\infty i^2\, t^{4(i-1)} \biggr]
\\
&= 2(1 - t^2)^4 \sum_{i=1}^\infty \biggl[ 2i(i+1)\, t^{2(2i-1)} 
+ i^2\, t^{4(i-1)} \biggr] 
= \frac{2(1 - t^2)}{1 + t^2}
\biggl[1 + \frac{2t^2}{(1 + t^2)^2} \biggr],
\end{align*}
after some calculation. So the purity of the first excited mode is
\[
\Pi_{\fs,x} 
= \frac{1 - t^2}{1 + t^2} \biggl[ 1 + \frac{2t^2}{(1 + t^2)^2} \biggr]
= \Pi_{\gs,1} \biggl[ 1 + \frac{2t^2}{(1 + t^2)^2} \biggr]
= \frac{2\sqrt{\om\mu}}{\om + \mu} \biggl(
1 + \frac{1}{2} \Bigl( \frac{\om - \mu}{\om + \mu} \Bigr)^2 \biggr).
\]
Since the other two modes contribute with two ground state factors,
the total purity can be written as
$\Pi_\fs = \Pi_{\fs,x} \Pi_{\gs,y} \Pi_{\gs,z}$. For the purity
parameter, one obtains finally
$$
s_\gs = 1 - \biggl( \frac{1 - t^2}{1 + t^2} \biggr)^3  \word{and}
s_\fs = 1 - \biggl( \frac{1 - t^2}{1 + t^2} \biggr)^3
\biggl[ 1 + \frac{2t^2}{(1 + t^2)^2} \biggr]
= s_\gs - \frac{2t^2(1 - t^2)^3}{(1 + t^2)^5}.
$$
In conclusion, $s_\fs\leq s_\gs$.

\smallskip

At long last, we may go back to Moshinsky's starting point, the
assessment of electron correlation, only now for the \textit{excited}
state. The Hartree--Fock approximation for the relevant mode, in view
of~\eqref{eq:rhodos-fs}, is of the form
\begin{align*}
W_\HF(r_1,r_2;p_1,p_2) &= \frac{1}{2} \bigl[
W_{00}(r_1;p_1) W_{11}(r_2;p_2) - W_{01}(r_1;p_1) W_{10}(r_2;p_2)
\\
&\qquad
- W_{10}(r_1;p_1) W_{01}(r_2;p_2) + W_{11}(r_1;p_1) W_{00}(r_2;p_2)
\bigr],
\\
\text{where }
W_{00}(r;p) &= \frac{1}{\pi} e^{-\eta r^2 - p^2/\eta},  \qquad
W_{11}(r;p) = \frac{2}{\pi} e^{-\eta r^2 - p^2/\eta}
(\eta r^2 + p^2/\eta - \half),
\end{align*}
with their corresponding interferences. Remember that 
$\int W_{ij} \,dr\,dp = \dl_{ij}$. In intracule-extra\-cule 
coordinates:
\[
W_\HF(R,r;P,p)
= \frac{2}{\pi^2} \bigl( \eta r^2 + p^2/\eta - \half \bigr)
e^{-\eta R^2 - P^2/\eta - \eta r^2 - p^2/\eta}.
\]

The parameter $\eta$ is determined by minimization. The mean value of
the energy predicted by this function is:
\begin{align*}
E_\HF &= \frac{1}{2} \int (p^2 + \om^2 r^2) [W_{00}(r;p) + W_{11}(r;p)] 
\,dr \,dp - \frac{\dl}{4} \int (r_1 - r_2)^2 \, W_\HF(1,2) \,d1 \,d2
\\
&= \biggl( \eta + \frac{\om^2}{\eta} \biggr) - \frac{3\dl}{4\eta}
= \eta + \frac{\om^2 + 3 \mu^2}{4\eta}.
\end{align*}
The minimum $dE/d\eta = 0$ occurs when
$\eta = \half\sqrt{\om^2 + 3\mu^2}$. Therefore, the energy predicted
by Hartree--Fock is $\sqrt{\om^2 + 3\mu^2}$. Thus, the ``correlation
energy'' for the lowest excited state of harmonium is:
\[
E_{\mathrm{c},\fs} = E_\fs - E_\HF
= \frac{3\om + 5\mu}{2} - \sqrt{\om^2 + 3\mu^2}
- 2\sqrt{(\om^2 + \mu^2)/2} \sim -\frac{7}{64}\,\frac{\dl^2}{\om^3}\,.
\]
Thus, the relative correlation energies are 
\[
\E_\fs := \frac{|E_{\mathrm{c},\fs}|}{E_\fs}
\sim \frac{7}{256}\,\frac{\dl^2}{\om^4}  \word{and}
\E_\gs := \frac{|E_{\mathrm{c},\gs}|}{E_\gs}
\sim \frac{1}{32}\,\frac{\dl^2}{\om^4} \,.
\]
Both quantities are related by a factor of $7/8$. For this 
approximation, as one would have
expected, $\E_\fs \leq \E_\gs$.

\begin{figure}[ht] 
\centering
\begin{minipage}[t]{.48\textwidth}
\begin{center}
\includegraphics[width=7.8cm]{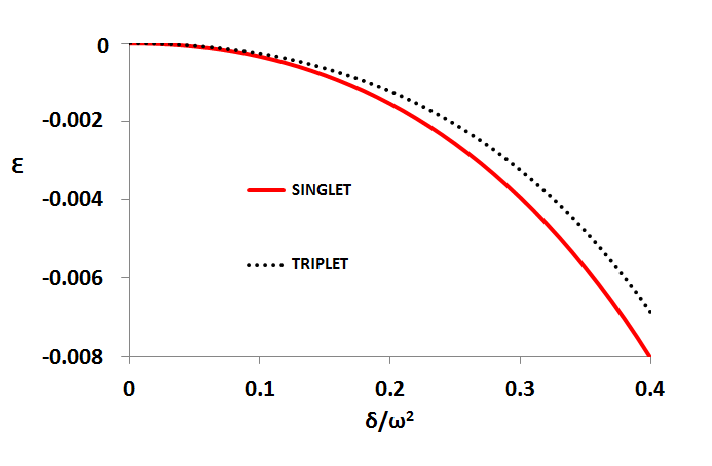} 
\end{center}
\end{minipage}
\hfill
\begin{minipage}[t]{.48\textwidth}
\begin{center}
\includegraphics[width=7.8cm]{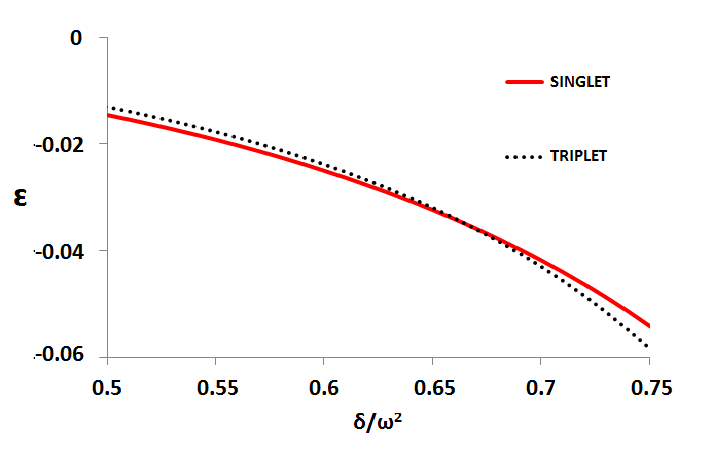}  
\end{center}
\end{minipage}
\hfill \caption{\textit{Relative} correlation energy of the singlet
and of the triplet excited mode. As expected, the relative correlation
energy for the singlet is greater than for the triplet for small
values of the coupling. At~$\dl/\om^2 \sim 0.67$ the order is
inverted.}
\label{graf:correlacion}
\end{figure}

Figure~\ref{graf:correlacion} shows the exact dependence of the
\textit{relative} correlation energy for both systems as a function
of~$\dl/\om^2$. The relative correlation energy for the singlet is
greater than for the triplet, just as the purity parameter for the
singlet is greater than the one for the triplet. At~$\dl/\om^2 = 0.67$
the relation between these two quantities changes and the relative
correlation energy for the triplet is greater than for the singlet.
Note however that the entropy depends only the behavior of the
occupation numbers, while the correlation energy has to do with the
natural orbitals as well. Such a nice proportionality
as~\eqref{eq:capeat-qui-potest} fails for the triplet state.

\begin{figure}[ht] 
\centering
\includegraphics[width=9cm]{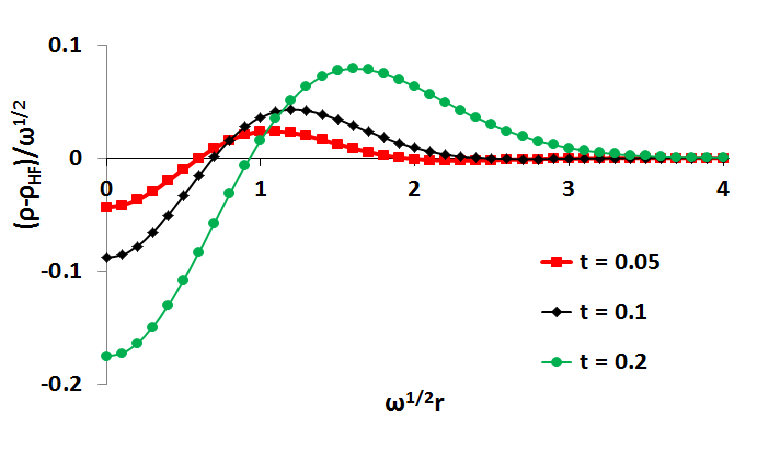}
\caption{Moshinsky's hole for the triplet:
$(\rho(r) - \rho_\HF(r))/\om^{1/2}$ as a function of $\om^{1/2} r$.}
\label{graf:MoshPit}
\end{figure}

Finally, Figure~\ref{graf:MoshPit} shows the difference between the
exact profile 1-density and the Hartree--Fock profile 1-density for
the harmonium triplet,
$\rho_\HF(r) := \int W_{\HF}(r,r_2;p_1,p_2) \,dp_1\,dr_2\,dp_2$. This
description goes back to the Coulson--Neilson classic
paper~\cite{CoulsonN61} on the helium Coulomb system. The
``Moshinsky's hole'' observed in the neighborhood of $r = 0$
graphically shows the Hartree--Fock underestimation of the mean
distance between the fermions, for the excited configuration of
harmonium as well.

\section{Conclusion}
\label{sec:conc}

From the very beginning of quantum mechanics, the fundamental state of
harmonium has provided a useful playground for learning about such
questions as correlation energy, entanglement or hole entropy
(including black hole entropy). Here, for the first time, we rather
exhaustively analyze the (spin triplet) first excited configuration of
harmonium, particularly the behaviour of its occupation numbers and
natural orbitals. This is a different chemical species altogether, due
to the antisymmetric character of the orbital wave function. When
exactly reconstructing \textit{\`a la} L\"owdin--Shull--Kutzelnigg the
two-body density as a functional of the one-body density, instead of
the sign dilemma (already solved by two of us) for the lowest-energy
state, we find, as expected on general grounds, an ambiguity in the
choice of natural orbitals.

Also as expected, in the triplet case the first occupation number
plays a more dominant role than for the singlet, up to fairly high
values of the coupling parameter, $t\lesssim 0.4$. Thus, within this
range, modeling the excited configuration as a Hartree--Fock state
introduces a lower error than doing so for the ground state. In
parallel, the linear entropy of the first excited configuration is
lower than that of the ground state, and the relative correlation
energy for the excited state stays below that of the ground state for
such values of the coupling. The order reverses at higher values of
$t$.

\subsection*{Acknowledgments}

JMGB thanks the Zentrum f\"ur interdisziplin\"are Forschung (ZiF) at
Bielefeld, in whose welcoming atmosphere this paper received its
finishing touches. CMBR and JMGB are grateful to Andr\'es F.
Reyes-Lega for an illuminating discussion. JCV thanks the Departamento
de F\'isica Te\'orica of the Universidad de Zaragoza for warm
hospitality.

CLBR and JMGB have been supported by grant FPA2009--09638 of Spain's
central government. CLBR thanks Banco Santander for support. JMGB owes
to ZiF for support, as~well. JCV acknowledges support from the
Direcci\'on General de Investigaci\'on e Innovaci\'on of Aragon's
regional government, and from the Vicerrector\'ia de Investigaci\'on 
of the University of Costa Rica.

Last, but not least, we thank the referee for very helpful criticism,
questions and suggestions, leading to an improved presentation.

\end{document}